\documentstyle[12pt]{article}
\def\bc{\begin{center}}                \def\ec{\end{center}}
\def\be{\begin{equation}}              \def\ee{\end{equation}}
\def\bear{\begin{eqnarray}}            \def\eear{\end{eqnarray}}
\def\bt{\begin{tabular}}               \def\et{\end{tabular}}
\def\la{\langle}     \def\ra{\rangle}   \def\dg{\dagger}
\def\ci{\cite}       \def\lb{\label}     \def\ld{\ldots}
      \def\vs{\vspace}   \def\pr{\prime}

  \def\rar{\rightarrow}
       \def\pr{\prime}

\def\a{\alpha}      \def\b{\beta}         
\def\D{\Delta}         
\def\l{\lambda}    \def\L{\Lambda}    \def\s{\sigma}    \def\z{\zeta}

\begin{document}

\begin{flushright}  {\small  Preprint INRNE-TH-98/2 (June 1998)\\
                                        E-print quant-ph/9809026\\
                                   J. Phys. A {\bf 31} (1998) .....}
\end{flushright}
\vspace{7mm}

\begin{center}
{\Large \bf Characteristic Uncertainty Relations }\\[5mm]

             {\large  D.A. Trifonov  and  S.G. Donev }\\[2mm]
           {\it Institute for Nuclear Research and Nuclear Energetics\\
            72 Tzarigradsko chauss\'ee, Sofia, Bulgaria }
\end{center} \vs{1cm}

\begin{minipage}{14cm}
{\small \bf Abstract}.
 {\small
 New uncertainty relations  for $n$ observables are established. The
 relations take the invariant form of inequalities between the
 characteristic coefficients of order $r$, $r=1,2,\ld,n$, of the
 uncertainty matrix and the matrix of mean commutators of the observables.
 It is shown that the second and the third order characteristic
 inequalities for the three generators of $SU(1,1)$ and $SU(2)$ are
 minimized in the corresponding group-related coherent states with maximal
 symmetry.}
\end{minipage}
\vs{1.2cm}

{\large \bf 1.\, Introduction}
\vs{5mm}

The uncertainty relations (UR) are basic nonclassical features of quantum
theory. In the last decades they have been extensively used in quantum
optics for constructing the so called nonclassical states \ci{Walls}.
In 1927 Heisenberg \ci{H} formulated the uncertainty principle as the
impossibility to determine simultaneously the position $q$ and momentum
$p$ of a particle with an accuracy higher than the Plank constant $\hbar$:
the product of the uncertainties $\D p,\,\,\D q$ in $p$ and $q$ should
not be less than $\hbar$, $\D p\D q \sim \hbar$. It was Weyl \ci{H} who
proved the "Heisenberg uncertainty relation" for $p$ and $q$
($\D^2p\D^2q\geq \hbar^2/4$) and Robertson \ci{Rob1} who extended the
latter to arbitrary two quantum observables (Hermitian operators) $X$ and
$Y$,

\be\lb{hur}   
\D^2X\D^2Y \geq \frac 14\left|\la [X,Y]\ra\right|^2,
\ee
where $[X,Y]$ is the commutator of $X$ and $Y$, $\la X\ra$ is the
average value of $X$ and $\D^2X$ is the variance of $X$. Nevertheless, the
inequality (\ref{hur}) is referred to as Heisenberg UR for two observables.
A more precise inequality for two observables was established by
Schr\"odinger \ci{Sch},

\be\lb{sur}    
\D^2X\D^2Y - (\D XY)^2 \geq \frac 14\left|\la
[X,Y]\ra\right|^2,
\ee
where $\D XY = \la XY + YX \ra/2 - \la X\ra\la Y\ra$ is the covariance of
$X$ and $Y$.
The coherent states (CS) \ci{KlSk}
and squeezed states \ci{Walls} of the one mode radiation field
widely discussed in the literature are pure
quantum states in which the inequalities (\ref{hur}) and/or (\ref{sur})
for the two canonical observables $p$ and $q$ ($[p,q]=-i\hbar$) are
minimized \ci{KlSk,DKM,T93}. The minimization of the Schr\"odinger UR for
two observables was considered in \ci{T94}, where the minimizing states
for two generators of the groups $SU(1,1)$ and $SU(2)$ have also been
constructed and discussed.

An important advantage of the Schr\"odinger formulation of
the uncertainty principle is the invariance of the equality in
(\ref{sur}) under linear nondegenerate transformations of $X$ and $Y$
\ci{T96a,T97a}, in particular under linear canonical transformations of
$p$ and $q$ \ci{DKM,SCB,T97a}.  Robertson \ci{Rob2} extended UR (\ref{hur})
to arbitrary $n$ observables $X_j$, $j = 1,2,\ld$, in the invariant form

\be\lb{rur}    
\det\s(\vec{X};\rho) \geq \det C(\vec{X};\rho),
\ee
where $\s(\vec{X};\rho)$ is the uncertainty (the dispersion or the
covariance) matrix of $n$ observables in the (generally mixed) state
$\rho$, $\s_{jk} = \la X_jX_k+X_kX_j\ra/2 - \la X_j\ra\la X_k\ra$, and
$C(\vec{X};\rho)$ is the $n\times n$ matrix of mean values of
the commutators $[X_j,X_k]$ times a factor $1/2i$, $\quad C_{jk} =
(-i/2)\la [X_j,X_k]\ra$. The uncertainty matrix $\s(\vec{X})$ is
symmetric, and the mean commutator matrix $C(\vec{X})$ is antisymmetric.
The other important advantage of Robertson UR is in its free integer
parameter $n$. This enables one to treat more complicated algebras (of
observables) with any finite dimension.

The aim of this letter is to establish new series of uncertainty
relations, which we call the {\it characteristic UR}.
The Robertson UR appears as one of the family of characteristic UR.

The idea is to consider the matrices $\s$ and $C$ as matrices of
linear maps in the $n$ dimensional vector space $E_n$, spanned by the
operators $X_j$, $j = 1,2,\ld,n$. The quantities $\det\s$ and $\det
C$, which enter the Robertson UR,  appear as characteristic coefficients
$C^{(n)}_n(\s)$ and $C^{(n)}_n(C)$ of the characteristic polynomials of
$\s$ and $C$ and the natural question arises whether there are inequalities
which relates the other characteristic coefficients $C^{(n)}_r(\s)$ and
$C^{(n)}_r(C)$, $r=0,1,\ld,n$.  The answer turned out to be positive and
the corresponding characteristic inequalities are established below.
\vs{1.2cm}

{\large \bf 2.\, Characteristic uncertainty relations}
\vs{5mm}

In order to extend the Robertson UR to the other characteristic
coefficients let us first recall \ci{T97a} the transformation properties
of the dispersion matrix $\s(\vec{X})$ and the matrix of mean commutators
$C(\vec{X})$ under linear transformations of the operators $X_j$,

\be\lb{X->X'}   
X_j \rar X_j^\pr =\l_{jk}X_k,\quad {\rm or\,\, in\,\, matrix\,\, form}
\quad \vec{X}^\pr = \L \vec{X}.
\ee
Using the definition of $\s$ and $C$ we easily obtain

\be\lb{trlaw}   
\s(\vec{X}^\pr) \equiv \s^\pr =
\L\s\L^{\rm T},\qquad C^\pr\equiv C(\vec{X}^\pr) = \L C \L^T,
\ee
where $\L^{\rm T}$ is the transposed of $\L$. If the transformation matrix
is real and nonsingular then the new $n$ operators $X_j^\pr$ are again
Hermitian and $\s^\pr$ is their dispersion matrix.

The transformation law (\ref{trlaw}) ensures the invariance of the
equality in (\ref{rur}) under nonsingular linear transformations of
observables (\ref{X->X'}),\,\, $\L \in GL(n,R)$.  If the operators $X_j$
close a Lie algebra $L$, then the equality in the Robertson UR is invariant
under the transformations of the group Aut$(L)$ of automorphisms of $L$.
It is curious that the equality in (\ref{rur}) is invariant under a wide
class of nonlinear state dependent transformations of $X_j$. Such are the
transformations (\ref{trlaw}) with $\L = \s$ or $\L = C$ in cases of $\det
C > 0$.

With the aim of establishing new uncertainty inequalities let us now
consider the characteristic equations for matrices $\s$ and $C$ ($\l$ and
$\mu$ are parameters),

\be\lb{chareqn}   
0 = \det(\s - \l) = \sum_{r=0}^{n} C^{(n)}_r(\s)(-\l)^{n-r},\qquad
0= \det(C - \mu) = \sum_{r=0}^{n} C^{(n)}_r(C)(-\mu)^{n-r}.
\ee
These equations are invariant under similarity transformations $\s \rar
\L\s\L^{-1}$, $C\rar \L C\L^{-1}$. The invariant coefficients
$C^{(n)}_r(\s)$\,\, ($C^{(n)}_r(C)$), $r=0,1,\ld,n$, in (\ref{chareqn})
are called the {\it characteristic coefficients} of $\s$\,\, ($C$)
\ci{Gantmaher}.  If one treats $\s$ and $C$ as linear maps in an $n$
dimensional vector space $E$ ($E\ni \vec{y}=\s \vec{x}$, $\vec{x}\in E$),
then $\s^\pr =\L\s\L^{-1}$ and $C^\pr =\L C\L^{-1}$ represent the same
maps in the new basis of $E$ (related to the old one by means of matrix
$(\L^{\rm T})^{-1}$).  The characteristic coefficients $C^{(n)}_r(\phi)$
of a matrix $\phi$ are equal \ci{Gantmaher} to the sum of all principle
minors $M(i_1,\ldots,i_r;\phi)$ of order $r$,

\be\lb{C_r}   
C^{(n)}_r(\phi) = \sum_{1\leq
i_1\leq i_2\leq \ldots\leq n}\left| \matrix{\phi_{i_1i_1}
\phi_{i_1i_2}\ldots\phi_{i_1i_r}\cr \phi_{i_2i_1} \phi_{i_2i_2}\ldots
\phi_{i_2i_r}\cr \ldots       \ldots       \ldots \ldots     \cr
        \phi_{i_ri_1} \phi_{i_ri_2}\ldots \phi_{i_ri_r}}\right| \equiv
    \sum_{1\leq i_1\leq i_2\leq \ldots\leq n}\,M(i_1,\ldots,i_r;\phi).
\ee \vs{2mm}

\noindent
One has $C^{(n)}_0 = 1$, $C^{(n)}_1 = {\rm Tr}\phi = \sum \phi_{ii}$ and
$C^{(n)}_n = \det\phi$. For $n=3$ we have, for example, three principle
minors of order $2$, i.e.,

\be\lb{C_2}    
C^{(3)}_2(\phi) =
\left|\matrix{\phi_{11} \phi_{12}\cr \phi_{21}\phi_{22}}\right| +
\left|\matrix{\phi_{11} \phi_{13}\cr \phi_{31}\phi_{33}}\right| +
\left|\matrix{\phi_{22} \phi_{23}\cr \phi_{32}\phi_{33}}\right|,
\ee \vs{2mm}

\noindent
where $|\phi| \equiv \det\phi$ for a matrix $\phi$.
In these notations Robertson UR (\ref{rur}) reads
$C^{(n)}_n\left(\s(\vec{X})\right) \geq C^{(n)}_n\left(C(\vec{X})\right)$
for any quantum state.  We shall now show that similar inequalities hold
for the other characteristic coefficients of $\s$ and $C$, namely

\be\lb{cur}   
C^{(n)}_r(\s(\vec{X})) \geq C^{(n)}_r(C(\vec{X})),\qquad r =
1,2,\ldots,n.
\ee
The key observation to this aim is that the principle submatrices
$\s(X_{i_1},\ldots,X_{i_r};\rho)$, $i_1 < i_2 < \ld <i_r$,

\be\lb{subsig}   
\s(X_{i_1},\ldots,X_{i_r};\rho) = \left(
\matrix{\s_{i_1i_1} \s_{i_1i_2} \ld \s_{i_1,i_r}\cr
        \s_{i_2i_1} \s_{i_2i_2} \ld \s_{i_2i_r} \cr
        \ld         \ld         \ld             \cr
        \s_{i_ri_1} \s_{i_ri_2} \ld \s_{i_ri_r}}  \right), \quad
    M(i_1,\ld,i_r;\s) = \det\s(X_{i_1},\ld,X_{i_r}),
\ee
can be regarded as uncertainty matrix for $r$ observables
$X_{i_1},\ld , X_{i_r}$ with $C(X_{i_1},\ldots,X_{i_r};\rho)$ as the
corresponding mean commutator matrix.  Therefore the inequality
(\ref{rur}) holds for the principle minors as well:

\be\lb{M_r}    
\det\s(X_{i_1},\ldots,X_{i_r};\rho) \geq \det
C(X_{i_1},\ldots,X_{i_r};\rho).
\ee
It is worth recalling now that
$\s(\vec{X})$ and $C(\vec{X})$ are nonnegative definite \ci{T97a} and
therefore all of their principle minors are also nonnegative
\ci{Gantmaher},

\be\lb{M>0}   
M(i_1,\ldots,i_r;\s) \geq 0,\,\,\,\, M(i_1,\ldots,i_r;C) \geq 0.
\ee
{}From (\ref{M>0}) and (\ref{C_r}) we derive that the inequalities
(\ref{cur}) do hold.  We shall call these inequalities the {\it
characteristic UR} for $n$ observables (Hermitian operators).  The two
sides of these relations are invariant under similarity transformations
$\s\rar \L\s\L^{-1}$ and $C\rar \L C\L^{-1}$.  The transformed matrix
$\s^\pr=\L\s\L^{-1}$ can be considered as an uncertainty matrix for new
$n$ observables $X^\pr_j$ in the same state $\rho$ iff $\L^{-1} = \L^T$ as
is seen from (\ref{trlaw}).  Therefore the characteristic UR (\ref{cur})
are invariant under linear transformation of the observables with
orthogonal $\L$.

On the other hand there are trace class invariant coefficients, related to
 any $n\times n$ matrix, and   one can look for uncertainty inequalities
involving these coefficients for the physical matrices $\s$ and $C$.  A
series of such inequalities for positive definite $2N\times 2N$ dispersion
matrices for $2N$ observables are established in \ci{T97a} (the particular
case of canonical observables being considered in \ci{SCB}),

\be\lb{traceUR}  
{\rm Tr}(i\s(\vec{X},\rho) J)^{2k} \geq 2^{1-2k}\sum_{j=1}^{N}\left|
\langle[X^\pr_\nu,X^\pr_{N+\nu}]\rangle\right|^{2k},\quad k=1,2,\ld,\quad
J = \left(\matrix{0 \,\,\,\mbox{-1}\cr 1 \,\,\,\,\,0}\right)
\ee
where
$X^\pr_j = \L(\rho)_{jl}X_l$, $\L$ being the symplectic matrix which
diagonalizes $\s(\vec{X};\rho)$ \ci{T97a}.  The traces ${\rm
Tr}(i\s(\vec{X},\rho) J)^{2k}$ are invariant under symplectic
transformations $\L$. If the operators $X_j$ close a Lie algebra $L$ with
a Cartan-Killing tensor $g$ then ${\rm Tr}(\s(\vec{X},\rho) g)^{2k}$ is
invariant under the group of automorphisms of $L$.  At $k=1=N$  one has
${\rm Tr}(i\s(\vec{X},\rho) J)^{2k} = \det\s(\vec{X},\rho)$ and
(\ref{traceUR}) coincides with (\ref{sur}).  \vs{1.2cm}

{\large \bf 3.\, Minimization of the Characteristic UR}
\vs{5mm}

The minimization of inequalities (\ref{hur}) and (\ref{sur}) proved
useful in constructing states with interesting physical and mathematical
properties.  States which minimize a certain uncertainty relation will be
called minimum uncertainty states (MUS) or intelligent states
(see \ci{T94,T97a} and references therein).  States which minimize
(\ref{sur})  (or (\ref{hur})) are called Schr\"odinger (Heisenberg) MUS or
intelligent states (or correlated coherent states \ci{DKM}).  For any pair
of observables $X$, $Y$ the necessary and sufficient condition for a state
$|\Psi\ra$ to minimize Schr\"odinger UR (\ref{sur}) is $|\Psi\ra$ to be an
eigenstate of a complex combination of $X,\,Y$ \ci{T94},

\be\lb{mincond1}    
(uA+vA^\dg)|\Psi\ra = z|\Psi\ra,\quad A=X+iY,\quad u,v\in C.
\ee
{}For $X,\,Y$ being the quadratures $p,\,q$ of boson annihilation operator
$a$ Schr\"odinger MUS coincide \ci{T93} with standard (or canonical)
squeezed states \ci{Walls}. The family of Schr\"odinger MUS
$|z,u,v;k\ra$ for the two quadratures of the Weyl lowering operator $K_-$
for the $su(1,1)$ algebra was constructed in \ci{T94} using the analytic
representation of Barut and Girardello \ci{BG}.

The minimization of Robertson UR was studied in
\ci{T97a}. Robertson MUS exist for a broad class of physical systems. It
was shown \ci{T97a} that group-related CS with maximal symmetry for
semisimple Lie groups are Robertson MUS for the quadratures of Weyl
lowering operators. For an odd number $n$ of observables $X_i$ a necessary
and sufficient condition for a state $|\Psi\ra$ to minimize (\ref{rur}) is
$|\Psi\ra$ to be an eigenstate of a real linear combination of all
observables. This condition remains sufficient for even $n$, $n = 2N$, as
well.  For even $n$ an other sufficient condition is $|\Psi\ra$ to be an
eigenstate of $N$ {\it complex} combinations of $X_i$,

\be\lb{mincond2}    
(\b_{\a i}X_i) |\Psi\ra \equiv (u_{\a \b}A_{\b} +
v_{\a \b}A^{\dg}_\b )|\Psi\ra = z_\a |\Psi\ra,
\ee
where $i=1,\ld,n$, $\a ,\b   = 1,2,\ld,N$, $A_\a  = X_\a +{\rm i}
X_{\a +N}$ and summation over repeated indices is adopted. The above
conditions are satisfied by the N-mode canonical CS ($u=1$, $v=0$, $A_\a
= a_\a $); by canonical squeezed states ($uu^\dg - vv^\dg =1$, $A_\a  =
a_\a $) and by canonical even/odd CS ($u=1$, $v=0$, $A_\a  = a^2_\a $),
i.e.  these important in quantum optics states are Robertson MUS for the
quadratures of all $a_\a $ and $a^2_\a $ correspondingly \ci{T97a}.

Using the structure (\ref{C_r}) of the characteristic coefficients and the
nonnegativity of the principle minors involved one can easily establish
the following minimization conditions for (\ref{cur}): \vs{2mm}

{\it Proposition 1}. The $r^{\rm th}$ order
characteristic UR (\ref{cur}) is minimized in a state $|\Psi\ra$ if
$|\Psi\ra$ is a Robertson MUS for every set of $r$ observables
$X_{i_1},\, X_{i_2},\ld, X_{i_r}$, $1 \leq i_1 < i_2 < \ld <i_r \leq n$.
\vs{2mm}

The characteristic UR of maximal order $r=n$ is that of Robertson and its
minimization conditions were listed just above \ci{T97a}.

Now it is of principle importance to show that the $r^{\rm th}$ order
characteristic MUS for $r \leq n$ do exist. It is clear that the first
order characteristic UR is trivial since it reads $\s_{11} + \s_{22} +\ld
+\s_{nn} \geq 0$, where all variances $\s_{ii} \equiv \D^2X_i$ are
nonnegative in any state. Its minimization is also clear - it is minimized
in the common eigenstates of $X_i$ only.  The case of $r = n$ was examined
in \ci{T97a}. So we have to provide the minimization of $r^{\rm th}$ order
characteristic UR, $2\leq r < n$.  To this aim let us consider the family
of $su(1,1)$ algebra related CS $|z,u,v,w;k\ra$, which are eigenstates of
the general element of the algebra in the representations $D^+(k)$, $k =
1/4,3/4$ and $k=1/2,1,\ld$,

\be\lb{|zuvw>}   
(uK_- + vK_+ +wK_3)|z,u,v,w;k\ra = z|z,u,v,w;k\ra.
\ee
Here $K_{\pm}=K_1\pm {\rm i}K_2$ are Weyl raising and lowering operators
and $K_{1,2,3}$ are Hermitian generators of the group $SU(1,1)$.
The solution to eq. (\ref{|zuvw>}) was obtained in \ci{T96a,B97a,T97a},
the case $w=0$ being solved previously in \ci{T94}. The large family of
$|z,u,v,w;k\ra$ contains all $SU(1,1)$ group-related CS with symmetry (see
\ci{KlSk} and references therein), the Schr\"odinger $K_1$-$K_2$ MUS
$|z,u,v,w=0;k\ra\equiv |z,u,v;k\ra$ containing the group-related CS with
maximal symmetry \ci{T94}. In the limit $u=1$, $v=0=w$ the CS $|z;k\ra$ of
Barut and Girardello \ci{BG} are reproduced: $|z;k\ra =
|z,u\!=\!1,v\!=\!0,w\!=\!0;k\ra$.  In the analytic Barut-Girardello
representation (which can be shown to be valid for $k=1/4,3/4$ as well)
the operators $K_i$ act as differential operators,

\be\lb{BGrep}       
K_+ = \eta,\quad K_-=2k\frac{\rm d}{{\rm d}\eta} +\eta\frac{{\rm d}^2}{{\rm
d}\eta^2},\quad K_3 = k +\eta\frac{\rm d}{{\rm d}\eta},
\ee
and the states $|z,u,v,w:k\ra$ are represented (for $u\neq 0$) by
the analytic functions

\be\lb{Phi_z}       
\Phi_z(\eta;u,v,w) = N\, e^{c\eta}\,_1F_1(a,b;c_1\eta),
\ee
where $N$ is the normalization factor, $a=k+z/l$, $b=2k$, $c = -(w+l)/2u$,
$c_1 = l/u$, $l \equiv \sqrt{w^2-4uv}$ and $_1F_1(a,b,z)$ is the confluent
hypergeometric function (the Kummer function) \ci{Stegun}. The limit $l
=0$ can be easily taken in ({\ref{Phi_z}), and the more simple case
of $u=0$ should be treated separately \ci{T96a,B97a,T97a}.

For the three observables $K_{1,2,3}$ we have two nontrivial
characteristic UR, namely those for $r = 2$ and $r=n=3$ in (\ref{cur}).
The Robertson relation for the three operators $K_{1,2,3}$ is minimized in
$|z,u,v,w;k\ra$ for $v=u^*$ and real $w$ and $z$ \ci{T97a}.
Such normalized intelligent states are, for example, the $SU(1,1)$
group-related CS $|\zeta;k\ra$ with maximal symmetry,

$$|\zeta;k\ra = N\,\exp(\z K_+)|k,k\ra,\,\,\z \in C,\,\, |\z|<1. $$

These CS, which in the BG representation are represented by the analytic
function $\exp(\zeta\eta)$ of the variable $\eta$ \ci{T94}, obey eq.
(\ref{|zuvw>}) with $v^* = u$, arg$u = -$arg$\z$, $w =w^* = |u|(1/|\z| -
|\z|)$ and $z = z^* = k|u|(-|\z| +1/|\z|)$, with $u$ remaining arbitrary.
We are now ready to prove that $|\zeta;k\ra$ minimize the second order UR
as well.

According to proposition 1, a state $|\Psi\ra$ minimizes the second
order characteristic UR for the three observables $K_{1,2,3}$ if it is an
eigenstate of the three combinations $\b_1 K_1 + \b_2 K_2 = uK_- + v K_+ +
0 K_3$,\, $\b^\pr_1 K_1 + \b^\pr_3 K_3 = u^\pr K_- + v^\pr K_+ + w^\pr K_3$
and $\b^{\pr\pr}_2 K_2 + \b^{\pr\pr}_3 = u^{\pr\pr} K_- + v^{\pr\pr} K_+ +
w^{\pr\pr} K_3$, for some real or complex $\b_{1,2}, \b^\pr_{1,2,3}$ and
$\b^{\pr\pr}_{1,2,3}$,

\be\lb{mincond3}       
\left. \bt{ll}
$(\b_1 K_1 + \b_2 K_2)|\Psi\ra  = z|\Psi\ra$,&\\[1mm]
$(\b^\pr_1 K_1 + \b^\pr_3 K_3)|\Psi\ra  = z^\pr|\Psi\ra$,&\\[1mm]
$(\b^{\pr\pr}_2 K_2 + \b^{\pr\pr}_3 K_3)|\Psi\ra  = z^{\pr\pr}|\Psi\ra$.
\et \right \}
\ee  \vs{3mm}

\noindent
In the representation (\ref{BGrep}) the system (\ref{mincond3}) takes
the form  of second order differential equations. One can check that
the functions

\be\lb{C_2cur}          
f(\eta,m) = \eta^m e^{-\z \eta},
\ee
satisfy the system (\ref{mincond3}) for $m=0$ and for $m=1-2k$ if

\be\lb{beta's}         
\b_1 = i\b_2\frac{1-\z^2}{1+\z^2};\,\,\, \b^\pr_1 =
2\b^\pr_3\frac{\z}{1+\z^2};\,\,\,\b^{\pr\pr}_2 =
2i\b^{\pr\pr}_3\frac{\z}{1-\z^2},
\ee
the eigenvalues being $z = (k+m\z)2i\b_2\z^2/(1+\z^2),\,\,
z^\pr  = (k+m)\b^\pr_3(1-\z^2)/(1+\z^2),\,\,
z^{\pr\pr} = (k+m)\b^{\pr\pr}_3(1+\z^2)/(1-\z^2)$
(with $\b^{\pr\pr}_3,\,\b^\pr_3$ and $\b_2$ remaining arbitrary).
This proves that the group-related
CS $|\zeta;k\ra$ are $C_2^{(3)}$ and
$C_3^{(3)}$ characteristic MUS. Let us recall that the group $SU(1,1)$ has
important (in quantum optics and other fields of quantum theory)
one-mode ($k=1/4,3/4$) and two-mode ($k=1/2,1,\ld$) boson representations.
In the one mode case the CS $|\z;k=1/4\ra$ coincides with the canonical
squeezed vacuum \ci{Walls}.

In a similar manner, using the results of papers \ci{T94,T97a,B97a} for
example, one can establish that the $SU(2)$ group-related CS with maximal
symmetry (the Bloch CS) are $C^{(3)}_2$ and $C^{(3)}_3$ characteristic
MUS.

Thus the characteristic UR can be used for finer classification of
quantum states.  For a given algebra they all are within the large set of
eigenstates of general algebra element (algebraic CS \ci{T96a,B97a}).
\vs{1cm}

{\large \bf 4.\, Concluding remarks}
\vs{4mm}

On the abstract matrix level Robertson proved \ci{Rob2,Bellmann} that if $H
= S +{\rm i}K$ is nonnegative definite Hermitian matrix (where $S$ and $K$
are real) then $\det S \geq \det K$. Matrix $\s(\vec{X}) +{\rm i}
C(\vec{X})$ is Hermitian and nonnegative, therefore $\det \s \geq \det
C$. Using Robertson' result we have proved in the above that if the
combination $S+{\rm i}K$ of two real matrices $S$ and $K$ is nonnegative
and Hermitian then the characteristic coefficients of $S$ and $K$ obey the
inequalities

\be\lb{last}    
C^{(n)}_r(S) \geq C^{(n)}_r(K), \quad r=1,2,\ldots,n.
\ee

The importance of the characteristic coefficients of a matrix is surely
 beyond doubt. In differential geometry and differential topology of fibre
 bundles with connections they are widely
 used as generators of topological invariants of the bundles by means of
 the De Rham cohomology of the corresponding base spaces \ci{Greub}.
In gauge field theories they are also well known and appropriately used
 \ci{Frankel}. It is our belief that the characteristic inequalities
(\ref{last}) will be useful in the above described fields as well.
\vs{2mm}

{\bf Acknowledgement}. This work was partially supported by the Bulgarian
Science Foundation under Contract no. F-559.
\vs{3mm}


\begin{thebibliography}{99}

\bibitem{Walls} Walls D.F. 1983, Nature {\bf 306} 141;\\
            Loudon R. and Knight P. 1987, J. Mod. Opt. {\bf 34} 709.

\bibitem{H} Heisenberg W. 1927, Ztschr. Phys., bd. 43, s. 172
            [Uspehi Fiz. Nauk, {\bf 122}, No. 4, 657 (1977)];\\
            Weyl H. 1928, {\it Theory of
            groups and quantum mechanics} (N.Y., Dutton).

\bibitem{Rob1} Robertson H.P. 1929, Phys. Rev. {\bf 34} 163.

\bibitem{Sch} Schr\"odinger E. 1930,
        Sitz. der Preuss. Acad. Wiss. (Phys. Math.  Klasse) (Berlin,
        1930), s.  296.

\bibitem{KlSk} Klauder J.R. and Skagerstam B.S. 1985 {\it
    Coherent States} (W. Scientific, Singapore).

\bibitem{DKM} Dodonov V.V., Kurmyshev E.V. and Man'ko V.I. 1980, Phys.
    Lett.  A {\bf 79} 150. {\small For a review on uncertainty relations
       see V.V.  Dodonov and V.I.  Man'ko, Trudy FIAN, v.  183, p.1-70
       (1987) (Nauka, Moscow, 1987).}

\bibitem{T93} Trifonov D.A. 1993, J. Math. Phys. {\bf34} 100.

\bibitem{T94} Trifonov D.A. 1994,  J. Math. Phys. {\bf35} 2297;
        D.A. Trifonov, {\it Generalized intelligent states and $SU(1,1)$
        and $SU(2)$ squeezing}, Preprint INRNE-TH-93/4 (May 1993).

\bibitem{T96a} Trifonov D.A. 1996, \, {\it Algebraic coherent states and
         squeezing}, E-print quant-ph/9609017.

\bibitem{T97a} Trifonov D.A. 1997, J. Phys. A {\bf30} 5941 [e-print
         quant-ph/9701018].

\bibitem{SCB} Sudarshan E.C.G., Chiu C.B. and Bhamati G. 1995, Phys. Rev.
        A {\bf52} 43.

\bibitem{Rob2} Robertson H.P. 1934, Phys. Rev. {\bf 46} 794.

\bibitem{Gantmaher} F.R. Gantmaher 1975, {\it Teoria matrits} (Nauka,
       Moscow).

\bibitem{BG}  Barut A.O. and Girardello L. 1971, Commun. Math. Phys.
        {\bf 21} 41.

\bibitem{B97a} Brif C.,\, Int. J. Theor. Phys. 1997 {\bf 36} 1677
     [E-print quant-ph/9701003].

\bibitem{Stegun} {\it Handbook of mathematical  functions},  edited  by M.
      Abramowitz and I A  Stegun  ({\i National Bureau of Standards},
     1964) (Russian translation: Nauka, Moscow, 1979).

\bibitem{Bellmann} Bellmann R. 1960, {\it Introduction to matrix
       analysis} (McGraw-Hill, New York - Toronto - London), chapter 8.

\bibitem{Greub} Greub W., Halperin S. and Vanstone R. 1972 {\it
Connections, Curvature and Cohomology} (Academic Press, New York and
        London).

\bibitem{Frankel} Theodore Frankel 1997, {\it The Geometry of Physics}
       (Cambridge University Press).

\end{thebibliography}
\end{document}